\documentclass[twocolumn,showpacs,showkeys,amsmath,aps,prc]{revtex4}
\newcommand{\bm}{\bibitem}

\usepackage{supertabular}
\usepackage{color,graphicx}
\usepackage[bookmarksopen]{hyperref}

\def  \g    {\gamma}
\def  \G    {\Gamma}
\def  \l    {\lambda}
\def  \L    {\Lambda}
\def  \o    {\omega}

\def  \s    {\sigma}

\def  \p    {\pi}

\def  \m    {\mu}
\def  \n    {\nu}

\def  \t    {\tau}
\def  \v    {\vec}
\def  \f    {\frac}
\def  \lt   {\left}
\def  \rt   {\right}

\def  \ra   {\rightarrow}

\def  \dt   {\delta}
\def  \Dt   {\Delta}
\def  \ep   {\epsilon}

\def  \be   {\begin{equation}}
\def  \ee   {\end{equation}}
\def  \ba   {\begin{array}}
\def  \ea   {\end{array}}
\def  \bea  {\begin{eqnarray}}
\def  \eea  {\end{eqnarray}}
\def  \nn   {\nonumber}
\def  \bd   {\begin{displaymath}}
\def  \ed   {\end{displaymath}}
\def  \bse  {\begin{subequations}}
\def  \ese  {\end{subequations}}
\def  \bwt  {\begin{widetext}}
\def  \ewt  {\end{widetext}}

\def  \vp   {{\bf p}}
\def  \vs   {{\bf \sigma}}

\begin{document}

\title{$\rho$-$\omega$ mixing and spin dependent CSV potential}

\author{Subhrajyoti Biswas}
\email{subhrajyoti.biswas@saha.ac.in}
\author{ Pradip Roy}
\author{Abhee K. Dutt-Mazumder}
\affiliation{Saha Institute of Nuclear Physics,
1/AF Bidhannagar, Kolkata-700 064, INDIA}

\medskip

\begin{abstract}

We construct the charge symmetry violating (CSV) nucleon-nucleon 
potential induced by the $\rho^0$-$\o$ mixing due to the neutron-proton 
mass difference driven by the $NN$ loop. Analytical expression for for the
two-body CSV potential is presented containing both the central and non-
central $NN$ interaction. We show that the $\rho$$NN$ tensor interaction 
can significantly enhance the charge symmetry violating $NN$ interaction 
even if momentum dependent off-shell $\rho^0$-$\omega$ mixing amplitude 
is considered. It is also shown that the inclusion of form factors removes
the divergence arising out of the contact interaction. Consequently, we see
that the precise size of the computed scattering length difference depends on 
how the short range aspects of the CSV potential are treated.

\end{abstract}
\vspace{0.08 cm}
\date{\today}



\maketitle

\section{introduction}

Charge symmetry violation (CSV), in itself, is an interesting 
physical phenomenon. While the charge symmetry (CS) implies that 
the interaction between two neutrons or two protons are equal, but, in 
nature, this is found to be only approximately true. The violation of CS 
automatically violates charge independence (CI), however, the 
converse might not be always true \cite{Stevens65,Henley72,Downs67}.
It is possible to have CS even if the CI is violated which actually 
is a higher symmetry. The CSV, at the QCD level, is caused by the
splitting of $u$-$d$ quark masses.

Experimentally CSV can be observed at various levels. For instance, 
in $NN$ interaction, the effect of CSV is traditionally inferred from
the difference of the $pp$ and $nn$ scattering lengths in the $^1S_0$ 
state. The most recent scattering data \cite{Miller90,Howell98,Gonzalez99}
observes that the amount of CSV in the $^1S_0$ state is $\Dt a_{CSV} = 
a^N_{pp} - a^N_{nn} = 1.6 \pm 0.6 ~fm$, where the superscript $N$ 
indicates the `nuclear' effect obtained after the electromagnetic (EM) 
corrections. Other convincing evidence of CSV $NN$ interaction comes from 
the binding energy difference of mirror nuclei which is known as Okamoto-Nolen
-Schifer (ONS) anomaly \cite{Nolen73,Okamoto64,Garcia92}. The modern manifestation 
of CSV includes difference of neutron-proton form factors, hadronic correction 
to $g-2$ \cite{Miller06} and the observation of the decay of $\Psi^{\prime}
(3686) \ra (J/\Psi) \p^0 $ etc \cite{Miller06}.

In the present work we focus on the hadronic sector, and, in 
particular, we attempt to construct CSV potential for the $NN$ 
interaction in one boson exchange (OBE) model by invoking momentum 
dependent $\rho^0$-$\omega$ mixing. The fact that the neutron and 
proton masses are not degenerate, the various isospin pure resonant
states like $\rho^0$-$\omega$ or $\pi^0$-$\eta$ can, in reality, mix 
without violating any conservation principles dictated by other symmetries.
In particular, the $\rho^0$-$\omega$ mixing seems to be a viable mechanism
for the generation of significant amount of CSV \cite{Henley79,McNamee75,Coon77,
Blunden87}. The earlier construction of CSV potential involved on-shell mixing 
of the $\rho^0$ and $\omega$ meson states \cite{Blunden87}. A whole class of 
phenomenon including the difference of $nn$-$pp$ scattering length, binding energy 
difference of $^3He$-$^3H$, or ONS anomoly in general could be successfully 
explained via $\rho^0$-$\omega$ mixing \cite{Sidney87}.

However, in ref.\cite{Piekarewicz92} such a success was severely 
criticised on ground that the on-shell mixing amplitude differs quite
significantly as one extrapolates the results from the $\rho$ (or 
$\omega$) pole to the space-like region which is relevant for the 
construction of the CSV potential. Goldman, Henderson and Thomas 
\cite{Goldman92} showed the strong ${\bf q}^2$ dependence of 
$\rho^0$-$\omega$ mixing to CSV potential using a simple quark model. 
Similar results were reported in ref.\cite{Piekarewicz92,Krein93,Connell94}.
In ref. \cite{Goldman92,Piekarewicz92,Krein93,Connell94,Coon97,Hatsuda94}
it was shown that such momentum dependencies suppress the contribution
of $\rho^0$-$\omega$ mixing and also changes the sign of the mixing 
amplitude as one moves away from the $\rho$ and $\omega$ poles.

On the other hand Miller \cite{Miller06} and Coon {\em et al} 
\cite{Coon97} have advanced counter arguments that would restore the 
traditional role of $\rho^0$-$\omega$ mixing. The issue is still 
unresolved. Informative summaries of the controversial point of 
views can be found in refs. \cite{Miller95,Connell97,Coon00}. Subsequently,
several other calculations were also performed including QCD sum-rule 
\cite{Hatsuda94,Akdm00} with varied conclusions. 

Recently Machleidt and M\"{u}ther \cite{Machleidt01} have discussed 
various CSV mechanism to estimate the $^1S_0$ scattering length. 
Therefore, the issues, including $\rho^0$-$\omega$ mixing as the origin 
of the CSV force, seem to be quite open which provide part of the 
motivation of the present work. 

Here we revisit the problem of $\rho^0$-$\omega$ mixing and invoke
the mechanism adopted in \cite{Piekarewicz92} {\em i. e.} the mixing
is driven by the neutron-proton mass difference. Although the driving 
mechanism is same, the main difference of our work with that presented 
in ref.\cite{Piekarewicz92} resides in the treatment of the external 
legs. This is another source of CSV due to the non-degenerate 
nucleon mass. This, as we shall see, has serious consequence which even 
modify the central part of the CSV potential. We highlight the importance 
of $\rho NN$ tensor interaction and show how this can counter balance the 
weakening of the strength of CSV interaction even when one extrapolates 
\cite{Henley79,McNamee75,Coon77} the results from on-shell to off-shell. 

The paper is organised as follows. In section II we present the formalism
where three momentum dependent $\rho^0$-$\omega$ mixing amplitude is used
for the construction of CSV potential. The numerical results including the 
contributions of external legs and the $\rho NN$ tensor coupling to CSV
potential are discussed in section III. Finally, we summarize in section
IV.

\section{formalism}

To calculate the $\rho^0$-$\omega$ mixing amplitude we use the following 
vector meson-nucleon interactions:

\bse
\bea
\mathcal{L}_{\o NN} & = &  g_\o\bar{\Psi}\g_\m\Phi^\m_\o \Psi 
\label{olag}\\
\mathcal{L}_{\rho NN} & = &  g_\rho \bar{\Psi}\lt[\g_\m +\f{C_\rho}{2M} 
\sigma_{\m\n}\partial^\m \rt]{\bf \t} \cdot {\bf \Phi}^\n_\rho \Psi 
\label{rlag}
\eea
\ese

where $C_{\rho} = f_{\rho}/g_{\rho}$ is the ratio of vector to tensor 
couplings. $\Psi$ and $\Phi$  denote nucleon and meson fields respectively.
The tensor coupling of $\o$ is not included in the present calculation
because it is  negligible in comparison to the vector coupling. All the 
parameters used in the present calculation are taken from those given by
the Bonn group \cite{Machleidt87}.

\begin{figure}[htb]
\begin{center}
\includegraphics[scale=0.45,angle=0]{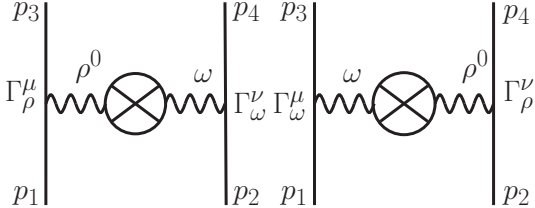}
\caption{Feynman diagrams for the mixing of isovector ($\rho^0$)-isoscalar 
($\omega$) mesons that contributes to the CSV $NN$ interaction. 
\label{fig1}}
\end{center}
\end{figure}

We now proceed to calculate the CSV potential using the Lagrangian
described above. The corresponding Feynman diagrams are shown in Fig.
\ref{fig1}. Here the CSV is represented by the crossed blobs 
$(\Pi_{\rho \omega}(q^2))$. This apart, the external legs, depending upon
whether we have proton or neutron, serve as another source of CSV as 
mentioned in the introduction due to their non-degenerate mass. Following
Fig.\ref{fig1} we write the matrix element as follows: 

\bea
\mathcal{M}^{NN}_{\rho \omega}(q)  &=&  [\bar{u}_N(p_3)\G_\rho^\mu u_N(p_1)]~
\Dt_{\mu \alpha}^\rho(q) \nn \\
         &\times & \Pi^{\alpha \beta}_{\rho \omega}(q^2)~
\Dt_{\beta \nu}^{\omega}(q)~ 
[\bar{u}_N(p_4)\tilde{\G}_{\omega}^{\nu} u_N(p_2)]. \label{ma0}
\eea

\begin{figure}[htb]
\begin{center}
\includegraphics[scale=0.35,angle=0]{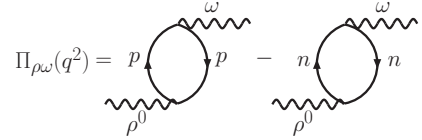}
\caption{ The mixing amplitude driven by the difference between 
proton and neutron loops due to $n$-$p$ mass difference.
\label{fig2}}
\end{center}
\end{figure}

The momentum space $NN$ potential $(V^{NN}_{\rho \omega}({\bf q}))$ can 
be obtained by taking the limit $q_0 \ra 0$ of $\mathcal{M}^{NN}_{\rho\o}(q)$.
Here $\G^\m_\o = g_\o\g^\m $, $\tilde{\G}^\n_\rho = g_\rho\lt[\g^\n 
+ \f{C_\rho}{2M}i\s^{\n\l}q_\l\rt]$ denote the vertex factors 
and $\Pi^{\mu \nu}_{\rho \omega}(q^2)$ is the mixing amplitude 
({\em i.e.} self-energy) driven by the difference between proton 
and neutron loops ({\em see} Fig.\ref{fig2}):

\bea 
\Pi^{\mu \nu}_{\rho \omega}(q^2) &=& \Pi^{\mu \nu (p)}_{\rho\o}(q^2)
-\Pi^{\mu \nu (n)}_{\rho\o}(q^2).
\label{se}
\eea 

The origin of the relative sign in the above equation is due to the different
signs involved in the coupling of $\o$ and $\rho^0$ with $p$ and 
$n$ ({\em see} Eqs. (\ref{olag})-(\ref{rlag})). It is to be noted that the 
$\rho NN$ vertex factor will have a relative sign depending upon whether 
it couples to $p$ or $n$.This sign flip has been included. 

The polarization tensor of $\rho^0$-$\o$ mixing due to $N\bar{N}$ 
excitations is calculated using standard Feynman rules:

\bea
i\Pi^{\m\n (N)}_{\rho\o}(q^2) &=& \int \f{d^4k}{(2\pi)^4}
Tr\lt[\G^\m_\o G_N(k) \tilde{\G}^\nu_\rho G_N(k+q) \rt]. \label{pol0}
\eea 

Here $G_N(k)$ is the usual Feynman propagator given by  

\bea
G_N(k) & = & \f{k\!\!\!/+M_N}{k^2-M^2_N+i\ep}. \label{np}
\eea

After performing the trace of Eq.(\ref{pol0}), one may write the 
polarization tensor as   
 
\bea
\Pi^{\m\n (N)}_{\rho\o}(q^2) &=& Q^{\m\n}
\lt[\Pi^{vv (N)}_{\rho\o}(q^2) + \Pi^{tv (N)}_{\rho\o}(q^2)\rt] 
\label{pol1}
\eea

where $Q^{\m\n} = (-g^{\m\n} + q^\m q^\n /q^2)$. The current conservation 
yields $q_\m \Pi^{\m\n}_{\rho\o}(q^2) = q_\n\Pi^{\m\n}_{\rho\o}(q^2) = 0$ 
as $q_\m Q^{\m\n} = q_\n Q^{\m\n} = 0$. From dimensional counting, it is 
clear that the integral in Eq.(\ref{pol0}) is ultraviolet divergent. We use
dimensional regularization \cite{Hooft73,Peskin95,Cheng06} to isolate 
the divergent parts of the integral in Eq.(\ref{pol0}) obtaining,

\bwt
\bse
\bea
\Pi^{vv(N)}_{\rho\o}(q^2) &=& -\f{g_\rho g_\o}{2\pi^2}\lt[\f{1}{6\ep}-
\f{\g}{6}- \int^1_0dx (1-x)x\ln\lt(\f{M^2_N - x(1-x)q^2}{\L^2} 
\rt)\rt] q^2\nn \\
\label{vv0}\\
\Pi^{tv(N)}_{\rho\o}(q^2) &=& -\f{g_\rho g_\o C_\rho}{8\pi^2}
\lt[\f{1}{\ep}-\g -\int^1_0dx \ln\lt(\f{M^2_N - x(1-x)q^2}{\L^2} 
\rt)\rt] q^2 \label{tv0},
\eea
\ese
\ewt

where $\L$ is an arbitrary renormalization constant; $\g$ is the 
Euler-Mascheroni constant. $\ep = 2 - D/2$ contains the singularity; 
$\ep \ra 0$ as $D \ra 4$. Since the mixing amplitude is the difference 
between proton and neutron loops contribution the divergent parts of
the above expressions cancel out yielding,

\bwt
\bse
\bea
\Pi^{vv}_{\rho\o}(q^2) &=&\Pi^{vv(p)}_{\rho\o}(q^2) - 
\Pi^{vv(n)}_{\rho\o}(q^2) = \f{g_\rho g_\o}{2\pi^2} \int^1_0 dx (1-x)x
\ln\lt(\f{M^2_p -x(1-x)q^2}{M^2_n -x(1-x)q^2} \rt) q^2, \nn \\
\label{vv1}\\
\Pi^{tv}_{\rho\o}(q^2) &=&\Pi^{tv(p)}_{\rho\o}(q^2)-
\Pi^{tv(n)}_{\rho\o}(q^2) = \f{g_\rho g_\o  C_\rho }{8\pi^2} \int^1_0 dx
\ln\lt(\f{M^2_p -x(1-x)q^2}{M^2_n -x(1-x)q^2} \rt) q^2,
\label{tv1}
\eea
\ese

The full mixing amplitude thus becomes,

\bea
\Pi{\rho\o}(q^2) &=& \Pi^{vv}_{\rho\o}({\bf q}) + \Pi^{tv}_{\rho\o}({\bf q})
=\f{g_\rho g_\o}{2\pi^2}q^2\int^1_0 
\lt((1-x)x + \f{C_\rho}{4}\rt) \ln\lt(\f{M^2_p -x(1-x)q^2}
{M^2_n -x(1-x)q^2} \rt) dx. \label{ma}
\eea
\ewt

Eq.(\ref{ma}) displays the four-momentum dependence of the $\rho^0$-$\o$ 
mixing amplitude in terms of three parameters $g_\rho$, $g_\o$ and 
$C_\rho$. We obtain $\Pi_{\rho\o}(m^2_\o) = -4314$ MeV$^2$  and 
$\Pi_{\rho\o}(m^2_\rho) = -4152$ MeV$^2$. These are within the 
limit of experimentally extracted values ($ \sim -4520 \pm 600$ MeV$^2$) 
\cite{Sidney87}. Upto now our results are same as that of 
ref.\cite{Piekarewicz92}. Note that most of the earlier efforts to understand
the role of $\rho^0$-$\o$ mixing in CSV potential was based on the assumption 
of constant on-shell value for the mixing amplitude 
\cite{McNamee75,Sidney87,Machleidt01}.

To calculate the CSV potential we have to use the mixing amplitude in 
spacelike region $(q_0 \ra 0)$. As a result the mixing amplitude becomes 
${\bf q}$ dependent {\em i.e.} $\Pi_{\rho\o}(0,{\bf q}) = \Pi_{\rho\o}({\bf q})$,
where we find

\be
\Pi_{\rho\o}({\bf q}) \simeq -\f{g_\rho g_\o }{12\pi^2}
(2+3C_\rho)\ln(M_p/M_n){\bf q}^2
\equiv -\mathcal{A} {\bf q}^2. \label{mix}
\ee

To calculate the CSV potential we take the non-relativistic (NR) limits of 
Eq.(\ref{ma0}). The relativistic energy $E_N$ is expanded in powers of ${\bf q}^2$ 
and ${\bf P}^2$ keeping the lowest order in ${\bf q}^2/M^2_N ({\bf P}^2/M^2_N)$
{\em i.e.} $E_N \simeq M_N + {\bf P}^2/2M_N + {\bf q}^2/8M_N$. Here,
${\bf P}=\f{1}{2}({\bf p}_2+{\bf p}_4)=-\f{1}{2}({\bf p}_1+ {\bf p}_3)$ 
is the average three momentum of the nucleon. The three momentum transfer 
is denoted by ${\bf q}=({\bf p}_1-{\bf p}_3 )=({\bf p}_4-{\bf p}_2)$ 
({\em see} Fig.\ref{fig1}). Also taking the NR limit of Dirac spinor and 
keeping terms $\mathcal{O}({\bf P}^2/M^2_N)$ and $\mathcal{O}({\bf q}^2/M^2_N)$
we obtain  

\bea
u_N(\vp_1) \simeq \lt(1-\f{{\bf P}^2}{8M^2_N}-\f{{\bf q}^2}{32M^2_N}\rt)
\left(\begin{array}{c}
1 \\
\f{\vs_1\cdot \lt({\bf P} + {\bf q}/2\rt)}{2M_{N}} \\
\end{array}
\right),
\label{sp}
\eea

where ${\bf \sigma}_{1(2)}$ is the nucleon spin. The relevant expressions 
which will be needed to construct the momentum space potential are the 
following:

\bse
\bea
\bar{u}_N({\bf p}_3)\g^0 u_N({\bf p}_1) &\simeq &  1+
\lt[ \f{{\bf P}^2}{4M^2_N} - \f{{\bf q}^2}{16M^2_N} + 
i\f{\vs_1\cdot({\bf q}\times{\bf P})}{4M^2_N} \rt], \nn \\
&  & \label{ex1}\\
\bar{u}_N({\bf p}_3)\v{\g} u_N({\bf p}_1) &\simeq &
\lt[\vs_1 \lt(\f{\vs_1 \cdot {\bf p}_1}{2M_N} \rt) +
\lt(\f{\vs_1 \cdot {\bf p}_3}{2M_N} \rt)\vs_1 \rt], \label{ex2} 
\eea
\bea
\bar{u}_N({\bf p}_4) \s_{l 0}q^l u_N({\bf p}_2) &\simeq &
i\lt(\f{{\bf q}^2}{2M^2_N}\rt),   \label{ex3} \\
\bar{u}_N({\bf p}_4) \s_{l k}q^l u_N({\bf p}_2) &\simeq &
- \lt(\vs_2\times{\bf q}\rt)_k ,  \label{ex4} 
\eea
\ese

where $(l, k)  = 1,2,3$. A straight forward calculation using Eqs.
(\ref{ex1})-(\ref{ex4}) and Eq.(\ref{ma0}) together with the NR
meson propagators leads to the momentum space CSV potential due to 
$\rho^0$-$\omega$ mixing: 

\bwt
\bea
V^{NN}_{\rho\o}({\bf q}) &=&-\f{g_\rho g_\o~\Pi_{\rho\o}({\bf q})}
{({\bf q}^2 + m^2_\rho)({\bf q}^2 + m^2_\o)} \nn \\
&\times&
\lt[T^+_3\lt\{ \lt( 1+\f{3{\bf P}^2}{2M^2_N}-\f{{\bf q}^2}{8M^2_N}  
- \f{{\bf q}^2}{4M^2_N} ({\bf \s}_1 \cdot {\bf \s}_2)
+ \f{3i}{2M^2_N}{\bf S}\cdot ({\bf q}\times {\bf P})
+\f{1}{4M^2_N}({\bf \s}_1\cdot {\bf q})({\bf \s}_2\cdot {\bf q})
+ \f{1}{M^2_N}({\bf \hat{q}}\cdot {\bf P})^2 \rt) \rt.\rt.
\nn \\ 
&-& \lt.\lt. \f{C_\rho}{2M} \lt(\f{{\bf q}^2}{2M_N} 
+ \f{{\bf q}^2}{2M_N}({\bf \s}_1 \cdot {\bf \s}_2)
- \f{2i}{M_N}{\bf S} \cdot ({\bf q}\times {\bf P}) 
- \f{1}{2M_N}({\bf \s}_1 \cdot {\bf q})({\bf \s}_2 
\cdot {\bf q})\rt) \rt\}\rt.
\nn \\
&-& T^-_3\lt. \f{C_\rho}{2M} \lt\{\lt(\f{{\bf q}^2}{2M}
-\f{{\bf q}^2}{2M}({\bf \s}_1\cdot{\bf \s}_2)
+\f{1}{2M}({\bf \s}_1\cdot{\bf q})({\bf \s}_2\cdot{\bf q})\rt)\f{\Dt M(1,2)}{M}
-\f{i}{M}({\bf \s}_1-{\bf \s}_2)\cdot({\bf q}\times{\bf P})\rt\}\rt]. 
\label{mpt}
\eea
\ewt

Here $T^\pm_3 = \t_3(1)\pm\t_3(2)$ and ${\bf S}=\f{1}{2}({\bf\s}_1+{\bf\s}_2)$ 
is the total spin of the interacting nucleon pair. We define $M=(M_n+M_p)/2$,
$\Dt M=(M_n-M_p)/2$ and $\Dt M(1,2)=-\Dt M(2,1) = \Dt M $.
The spin dependent parts of the momentum space potential as found in Eq.(\ref{mpt})
appear because of the contribution of the external legs shown in Fig.\ref{fig1}. 
On the other hand, $3{\bf P}^2/2M^2_N$ and $-{\bf q}^2/8M^2_N$ arise due to expansion 
of the relativistic energy $E_N$. 

Note that the potential derived in Eq.(\ref{mpt}) contains both class $(III)$ and
class $(IV)$ potentials, and both of these potentials break the charge symmetry of 
$NN$ interactions. The first part of this potential represents class $(III)$ $NN$ 
interaction which differentiates between $nn$ and $pp$ systems but vanishes for $np$ 
system. On the other hand, last part of Eq.(\ref{mpt}) is class $(IV)$ $NN$ interaction
which exists for $np$ system only. In the present paper, we focus on the class $(III)$ 
$NN$ potential.

From Eq.(\ref{mpt}) we extract a piece which, in coordinate space, gives rise to 
$\dt$-function potential. In momentum space it is given by 

\bea
\dt V^{NN}_{\rho\o} &=& g_\rho g_\o \mathcal{A}T^+_3
\lt[ \lt(\f{1+2C_\rho}{8M^2_N}\rt) + 
\lt(\f{1+C_\rho}{4M^2_N}\rt)({\bf \s}_1 \cdot {\bf \s}_2)\rt]. \nn \\
\label{dfp}
\eea   

The problem of contact term can be avoided by using form factors, for which
$g_i$ is repleced with $g_i({\bf q}^2)$.  

\bea
g_i \ra g_i({\bf q}^2)=g_i\lt(\f{\L^2_i-m^2_i}{\L^2_i+{\bf q}^2}\rt)
\label{ff}
\eea   

The cut-off parameters $\L_i$ governs the range of the suppression, which
can be directly related to the hadron size. The values of $\L_i$s 
are determined from the fit of the two-nucleon empirical data 
\cite{Machleidt87,Machleidt86}. 

The spin independent central part neglecting the contributions due to 
external legs and the $\rho NN$ tensor coupling, reduces to 

\bea
V^{0}_{\rho\o}({\bf q}) = -\f{g_\rho g_\o~\Pi_{\rho\o}({\bf q})T^+_3}
{({\bf q}^2 + m^2_\rho)({\bf q}^2 + m^2_\o)}, 
\label{vc}
\eea

which is same as obtained in ref.\cite{Piekarewicz92}. In coordinate
space, treating the on-shell mixing amplitude to be constant one obtains

\bea
V^0_{\rho\o}({\bf r}) = -\f{g_\rho g_\o}{4\pi}\f{\Pi_{\rho\o}(m^2_\o)T^+_3}
{m^2_\o - m^2_\rho} \lt[m_\rho Y_0(x_\rho) - m_\o Y_0(x_\o)\rt],\nn \\
\label{vcr}
\eea

where $Y_0(x_i) = e^{-x_i}/x_i$ and $x_i = m_ir, ~(i = \rho,\o)$. With form
factors Eq.(\ref{vcr}) reduces to 

\bwt
\bea
V^0_{\rho\o}({\bf r}) &=& -\f{g_\rho g_\o}{4\pi}\f{\Pi_{\rho\o}(m^2_\o)T^+_3}
{m^2_\o - m^2_\rho} 
\lt[
\lt\{\lt( \f{\L^2_\o-m^2_\o}{\L^2_\o-m^2_\rho} \rt) m_\rho Y_0(x_\rho)
-\lt( \f{\L^2_\rho-m^2_\rho}{\L^2_\rho-m^2_\o} \rt) m_\o Y_0(x_\o) \rt\} \rt. \nn \\
&+& \f{m^2_\o-m^2_\rho}{\L^2_\o-\L^2_\rho} \lt.
\lt\{ \lt( \f{\L^2_\o-m^2_\o}{\L^2_\rho-m^2_\o} \rt) \L_\rho Y_0(X_\rho) -
\lt( \f{\L^2_\rho-m^2_\rho}{\L^2_\o-m^2_\rho} \rt) \L_\o Y_0(X_\o) \rt\}
\rt], \label{vcrf}
\eea
\ewt

where $X_i = \L_i r$. Eq.(\ref{vcrf}) represents the CSV potential with constant
mixing amplitude neglecting the contribution of external legs. It is to be noted 
that in the linit $\L_{\rho,\o} \ra \infty$, Eq.(\ref{vcrf}) reduces to Eq.(\ref{vcr}).

If we include the contribution of the external legs  and $\rho NN$ 
tensor coupling, simplifies to the central part, 

\bea
V^{0NN}_{\rho\o}({\bf r}) &=& -\f{g_\rho g_\o}{4\pi} \mathcal{A}T^+_3 
\lt[\lt(\f{m^3_\rho Y_0(x_\rho) - m^3_\o Y_0(x_\o)}{m^2_\o - m^2_\rho}\rt)\rt. \nn \\
&+& \f{1+2C_\rho}{8M^2_N} \lt.
\lt(\f{m^5_\rho Y_0(x_\rho)- m^5_\o Y_0(x_\o)}{m^2_\o-m^2_\rho}\rt) \rt].
\label{vcnn}
\eea

In the above equation the first term in the bracket is same as one would
have obtained from Eq.(\ref{vc}) by taking the momentum dependent mixing
amplitude as in Eq.(\ref{mix}), while the second term contains the contribution
coming from the Dirac spinors of the external lines. The latter, clearly
involves $\rho NN$ vector and tensor interactions, and, as we shall see, 
the term containing the tensor coupling ($C_\rho $) is significantly 
larger compared to the vector interaction at distances below $0.75 fm$ or
so.

We leave out the coordinate space contact terms from Eq.(\ref{vcnn}) 
and Eq.(\ref{vrw}). We also drop the term $3{\bf P}^2/2M^2_N$ from Eq.(\ref{mpt})
while deriving the total coordinate space potential as it is not important in 
the present context. However, it should be noted that to fit the $^1S_0$ and $^3P_2$ 
phase shifts simultaneously this term is necessary as ${\bf P}^2$ gives the operator 
$\nabla^2_R$ in coordinate space. Moreover, we use the ${\bf q}^2$ dependent 
mixing amplitude instead of constant on-shell value and for this we consider 
terms upto $\mathcal{O}({\bf q}^2/M^2_N)$. Taking all this into consideration 
we obtain, after some algebraic manipulations, the coordinate space CSV 
potential as

\bea
V^{NN}_{\rho\o}({\bf r}) &=& -\f{g_\rho g_\o}{4\pi} \mathcal{A}T^+_3
\lt[\lt(\f{m^3_\rho Y_0(x_\rho) - m^3_\o Y_0(x_\o)}{m^2_\o - m^2_\rho}\rt) \rt. \nn \\
&+& \f{1 }{M^2_N} \lt.
\lt(\f{m^5_\rho V_{vv}(x_\rho) - m^5_\o V_{vv}(x_\o)}{m^2_\o - m^2_\rho}\rt) \rt. \nn \\
&+& \lt. \f{C_\rho  }{2M^2_N} \lt(\f{m^5_\rho V_{tv}(x_\rho) - m^5_\o 
V_{tv}(x_\o)}{m^2_\rho -m^2_\o}\rt)\rt]. \label{vrw}
\eea
 
The spin-spin, tensor and spin-orbit interaction terms are explicitly 
containted in $V_{vv}(x)$ and $V_{tv}(x)$ which are as follows:

\bse
\bea       
V_{vv}(x) &=& \f{1}{8}Y_0(x) + \f{1}{6}Y_0(x)({\bf \s}_1\cdot{\bf \s}_2) \nn \\
&-& \f{1}{12}Y_1(x)S_{12}(\hat{\bf r}) - \f{3}{2}Y_2(x){\bf L}\cdot {\bf S} \label{vvv}\\
V_{tv}(x) &=& \f{1}{2}Y_0(x) + \f{1}{3}Y_0(x)({\bf \s}_1\cdot{\bf \s}_2) \nn \\
&-& \f{1}{6}Y_1(x)S_{12}(\hat{\bf r}) - 2Y_2(x){\bf L}\cdot {\bf S} \label{vtv}.
\eea
\ese

where,

\bse
\bea
Y_1(x) &=& \lt(1 + \f{3}{x} + \f{3}{x^2} \rt)Y_0(x)\\
Y_2(x) &=& \lt(\f{1}{x} + \f{1}{x^2} \rt)Y_0(x) \\
S_{12}(\hat{\bf r}) &=& 3(\s_1\cdot\hat{\bf r})(\s_2\cdot\hat{\bf r})
- (\s_1\cdot\s_2)
\eea
\ese

The first part of Eq.(\ref{vrw}) represents the central part 
without contributions from external legs. In addition, the last two terms 
of Eq.(\ref{vrw}) are the contributions coming from the external nucleon legs
as discussed earlier. It is also to be noted that the central part also 
receives contributions due to the presence of the first terms in Eq.(\ref{vvv})
and Eq.(\ref{vtv}). The tensor contribution $(C_\rho)$ of $\rho$-meson is 
contained in the third term of Eq.(\ref{vrw}).

Eq.(\ref{vrw}) does not include form factors. It diverges near the core. This divergence
can be removed by incorporating form factors as in Eq.(\ref{ff}). Thus the complete CSV 
potential with form factors reduces to

\bwt
\bea
V^{NN}_{\rho\o}({\bf r}) &=& -\f{g_\rho g_\o}{4\pi}\f{\mathcal{A}T^+_3}{m^2_\o-m^2_\rho} 
\lt[
\lt\{ 
\lt(
\lt( \f{\L^2_\o-m^2_\o}{\L^2_\o-m^2_\rho} \rt) m^3_\rho Y_0(x_\rho)
-\lt( \f{\L^2_\rho-m^2_\rho}{\L^2_\rho-m^2_\o} \rt) m^3_\o Y_0(x_\o) 
\rt) \rt. \rt. \nn \\
&+& \f{1}{M^2_N}\lt.\lt.
\lt(
\lt( \f{\L^2_\o-m^2_\o}{\L^2_\o-m^2_\rho} \rt) m^5_\rho V_{vv}(x_\rho) -
\lt( \f{\L^2_\rho-m^2_\rho}{\L^2_\rho-m^2_\o} \rt) m^5_\o V_{vv}(x_\o) 
\rt) \rt. \rt. \nn \\
&+&\f{C_\rho}{M^2_N}\lt. \lt.
\lt(
\lt( \f{\L^2_\o-m^2_\o}{\L^2_\o-m^2_\rho} \rt) m^5_\rho V_{tv}(x_\rho) 
-\lt( \f{\L^2_\rho-m^2_\rho}{\L^2_\rho-m^2_\o} \rt) m^5_\o V_{tv}(x_\o) 
\rt) 
\rt\} \rt.\nn \\
&+& \lt(\f{m^2_\o-m^2_\rho}{\L^2_\o-\L^2_\rho}\rt)\lt.
\lt\{
\lt(
\lt( \f{\L^2_\o-m^2_\o}{\L^2_\rho-m^2_\o} \rt) \L^3_\rho Y_0(X_\rho) -
\lt( \f{\L^2_\rho-m^2_\rho}{\L^2_\o-m^2_\rho} \rt) \L^3_\o Y_0(X_\o) 
\rt) \rt. \rt.\nn \\
&+& \f{1}{M^2_N}\lt. \lt.
\lt(
\lt( \f{\L^2_\o-m^2_\o}{\L^2_\rho-m^2_\o} \rt) \L^5_\rho V_{vv}(X_\rho)-
\lt( \f{\L^2_\rho-m^2_\rho}{\L^2_\o-m^2_\rho} \rt) \L^5_\o V_{vv}(X_\o) 
\rt) \rt. \rt. \nn \\
&+& \f{C_\rho}{2M^2_N}\lt. \lt.
\lt(
\lt( \f{\L^2_\o-m^2_\o}{\L^2_\rho-m^2_\o} \rt) \L^5_\rho V_{tv}(X_\rho)-
\lt( \f{\L^2_\rho-m^2_\rho}{\L^2_\o-m^2_\rho} \rt) \L^5_\o V_{tv}(X_\o) 
\rt) \rt\}
\rt]. \label{vrwf}
\eea
\ewt
 
Note that the above equation contains the contribution of Eq.(\ref{dfp}). 
The CSV $NN$ potential given in Eq.(\ref{vrwf}) can be used to calculate the 
difference between $nn$ and $pp$ scattering lengths at $^1S_0$ state. The difference
between scattering lengths, $\Dt a = a_{pp}-a_{nn}$, and the difference between
CSV $nn$ and $pp$ potential, $\Dt V_{\rho\o} = V^{nn}_{\rho\o}-V^{pp}_{\rho\o}$,
are related by 

\bea
\Dt a = -a^2M\int^\infty_0 \Dt V_{\rho\o}~u^2_0(r)~dr
\eea    

where $a^2=a_{nn}a_{pp}$ and $u_0(r)$ is the zero energy wave function, 
normalized to approach $1-r/a$ as $r\ra\infty$ and $u(0)=0$. To calculate
$\Dt a$ we use the following zero energy wave function \cite{Mark90}:

\bea
u_0(r)&=&\lt[1-\f{r}{a}\rt]
-\lt[\g(1-\l)\f{r}{2}+(1+\l)\rt]\f{e^{-\g r}}{1+\l e^{-\g r}}, \nn \\
&&\label{zewf}
\eea

where $\l=(1-2r_0/a)^{-1/2}$, $\g=2(1+\l)/(r_0\l)$ and $r_0$ is the effective
range. In the present calculation we take $r_0=2.8$ fm.

\section{results}

\begin{figure}[htb]
\begin{center}
\resizebox{7.0cm}{6.5cm}{\includegraphics[]{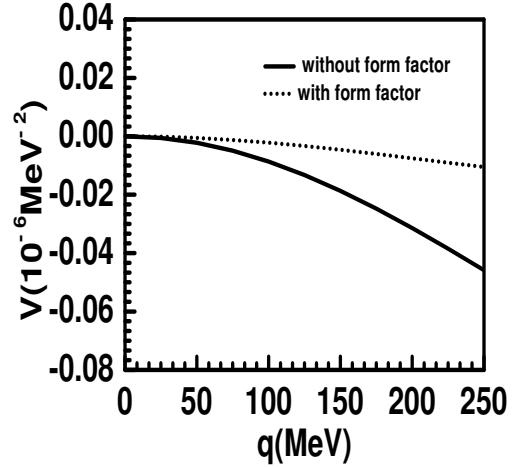}}
\caption{The $\rho^0$-$\o$ mixing contribution to the central part of 
the CSV $NN$ potential in momentum space (Eq.\ref{vc}) is presented here.
The dotted and solid curves present the three momentum dependent potential
with and without form factor, respectively.
\label{fig3}}
\end{center}
\end{figure}

In this section we present our results. First we show the momentum space 
central potential ({\em see} Eq.\ref{vc}) in Fig.\ref{fig3} considering the three
momentum dependent mixing amplitude. In this figure dotted curve represents
the central potential with the form factor in the space-like region. 
In contrast, the solid curve represents the same without the form factor 
\cite{Stevens65,Yalcin79}.

\begin{figure}[htb]
\begin{center}
\resizebox{7.0cm}{6.5cm}{\includegraphics[]{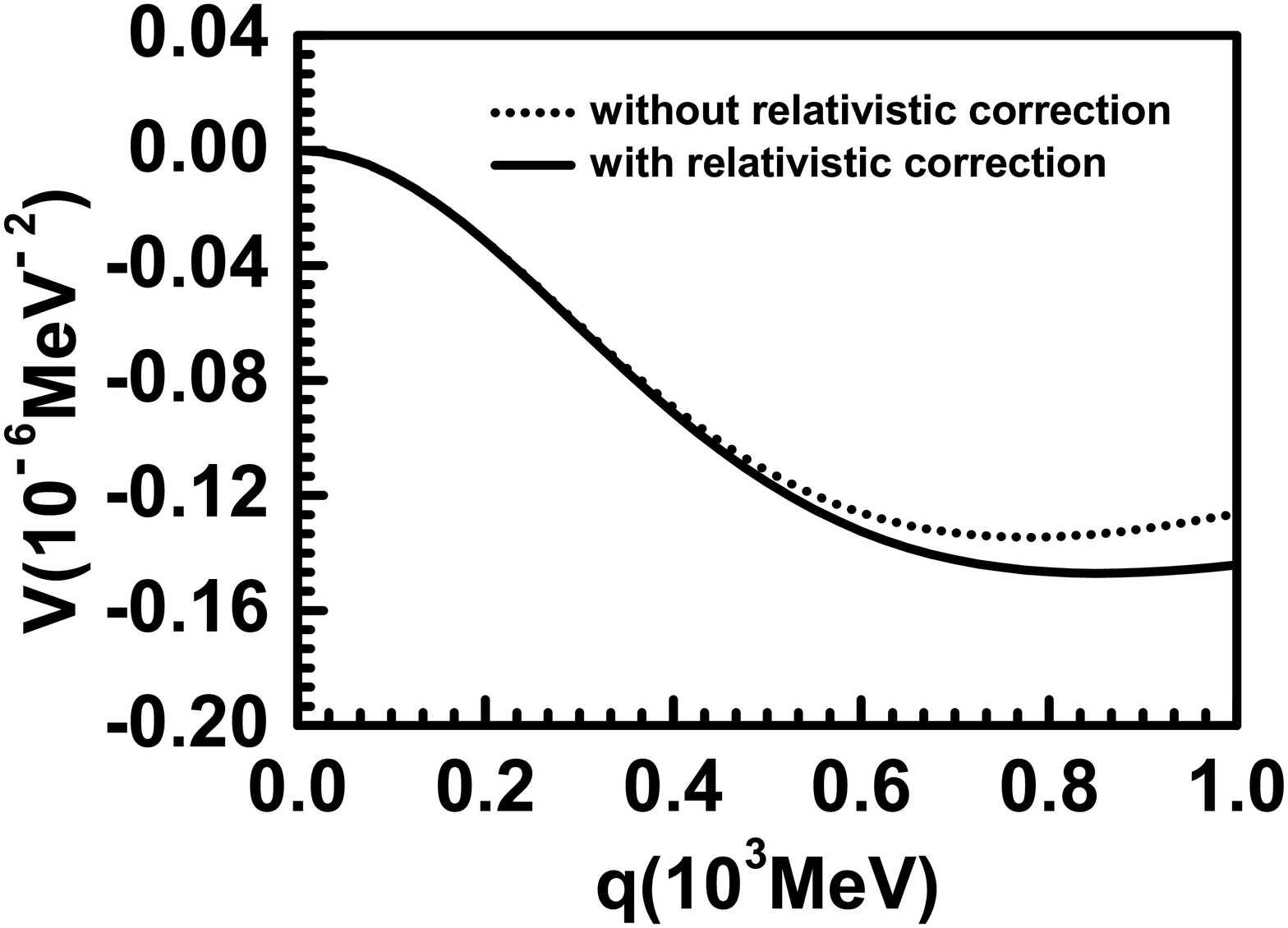}}
\caption{Central part of momentum space potential without
(dotted curve) and with (solid curve) the relativistic correction.
\label{fig4}}
\end{center}
\end{figure}

In Fig.\ref{fig4} the importance of the relativistic correction to the central
part potential in momentum space is displayed. This correction, as expected,
is marginal at low  momentum (below $|{\bf q}| \sim 500$ MeV) transfer. 
In the short distance regime {\em i.e.} near the core region, the relativistic 
correction becomes significant which is clearly seen in Fig.\ref{fig4}.

\begin{figure}[htb]
\begin{center}
\resizebox{7.0cm}{6.5cm}{\includegraphics[]{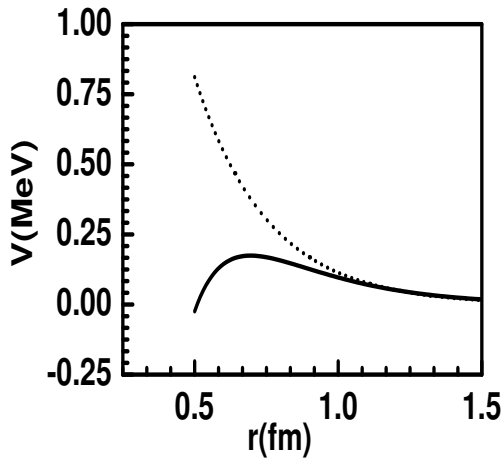}}
\caption{Central part of the coordinate space potential
without form factors. The potentials considering the constant 
on-shell mixing amplitude (dotted curve) and three momentum 
dependent mixing amplitude (solid curve) without external 
legs and $\rho NN$ tensor contributions.
\label{fig5}}
\end{center}
\end{figure}

\begin{figure}[htb]
\begin{center}
\resizebox{7.5cm}{7.0cm}{\includegraphics[]{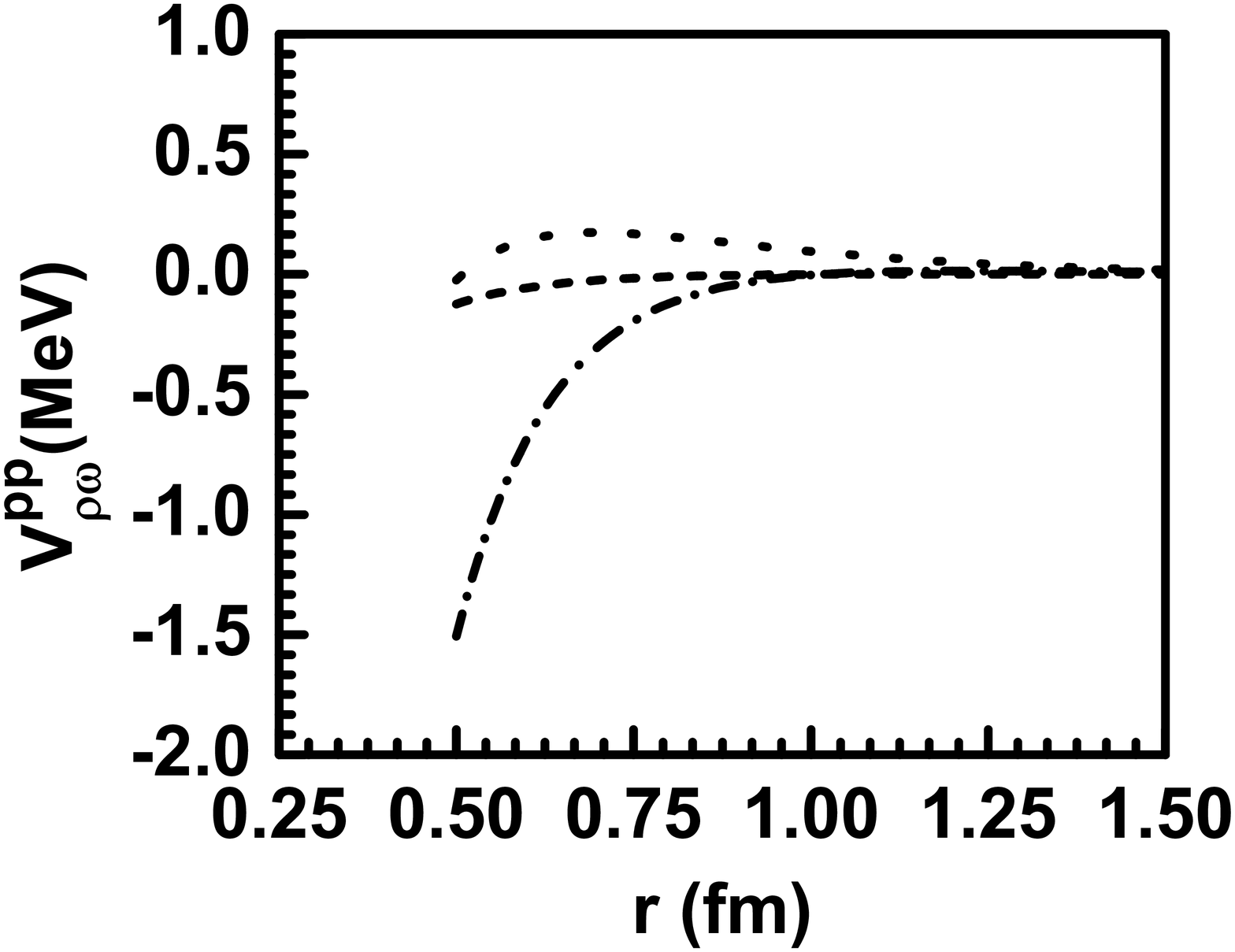}}
\caption{The dotted curve represents only the central part (Eq.\ref{vcnn}). 
The external legs and the $\rho NN$ tensor contributions to the central
potential are shown by dashed and dot-dashed curves, respectively. 
\label{fig:vcnn}} 
\end{center}
\end{figure}

In Fig.\ref{fig5} we show the central part of the potential due to both 
on-shell and off-shell mixing amplitudes. It is seen that the contribution 
of the off -shell $\rho^0$-$\o$ mixing amplitude to the $NN$ potential is 
opposite in sign relative to the contribution obtained from using the on-shell 
value. This, again, is consistent with the observation made in 
ref.\cite{Piekarewicz92}.

\begin{figure}[htb]
\begin{center}
\resizebox{7.5cm}{7.0cm}{\includegraphics[]{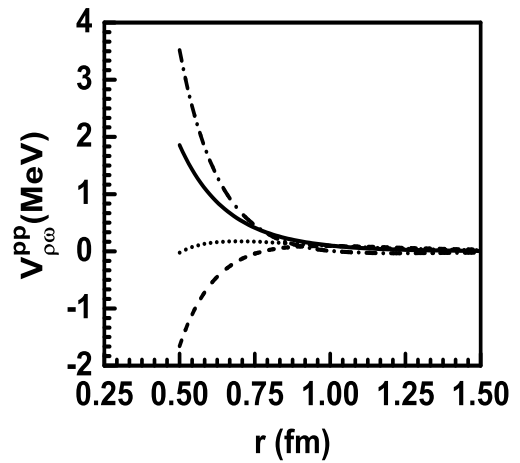}}
\caption{The coordinate space $V^{pp}_{\rho\o}$ potential without
form factors, at $^1S_0$ state. Different parts of CSV potential {\em i.e.} 
the central (dotted), central with external legs plus the $\rho NN$ tensor 
contribution (dashed) and the spin dependent (dashed-dotted) parts are presented 
here. The solid curve shows the total CSV potential. 
\label{fig6}} 
\end{center}
\end{figure}

The individual contribution of different parts of the central 
potential given in Eq.(\ref{vcnn}) is presented in Fig.\ref{fig:vcnn}. 
Clearly the contribution of $\rho NN$ tensor coupling to the CSV 
potential is found to be much larger than the contribution of the first part 
({\em i.e.} the central part without external legs and $\rho NN$ tensor
contribution). It is to be noted that $\rho NN$ tensor contribution is
present only when the external legs are taken into account.

The $^1S_0$ state CSV potential for  $pp$ system due to $\rho^0$-$\o$
mixing is shown in Fig.\ref{fig6}. The importance of the central part 
with relativistic correction (dashed curve) and tensor contribution (dashed-
dotted curve) are clearly revealed. The magnitude of the contribution of
tensor coupling is comparable with that of the central part with 
relativistic correction in the core region. On the other hand, magnitude 
of the contribution of tensor coupling is found to be much larger than the 
contribution of the central part (dotted curve) without relativistic correction
in the core region. In the dynamical region, it is seen that all the
contributions are comparable. The solid curve in this figure represents the 
total contribution together with the relativistic correction. 

\begin{figure}[*htb]
\begin{center}
\resizebox{7.5cm}{7.0cm}{\includegraphics[]{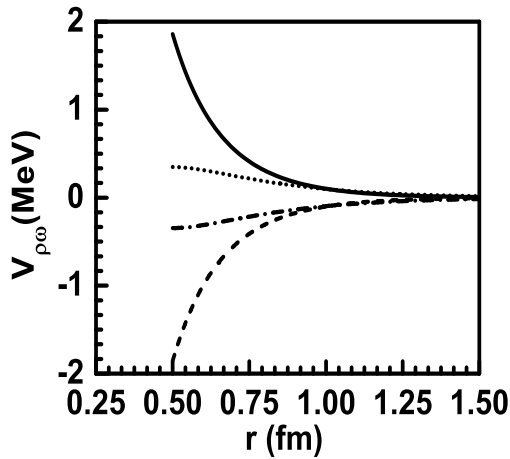}}
\caption{Total $^1S_0$ CSV potential without form factors for $pp$ and $nn$
systems are denoted by the Solid and the dashed curves, respectively. The same 
are presented by dotted and dashed-dotted curves without $\rho NN$ tensor contribution.
\label{fig7}}
\end{center}
\end{figure}

In Fig.\ref{fig7} we present the CSV potential at $^1S_0$ state both for
the $pp$ and $nn$ system. The solid and dashed curves, respectively, show 
the CSV potential for $pp$ and $nn$ system taking the contribution of tensor 
coupling of $\rho$-meson. The same are presented by dotted and dot-dashed
curves without considering the tensor coupling. 

\begin{figure}[*htb]
\begin{center}
\resizebox{8.5cm}{7.0cm}{\includegraphics[]{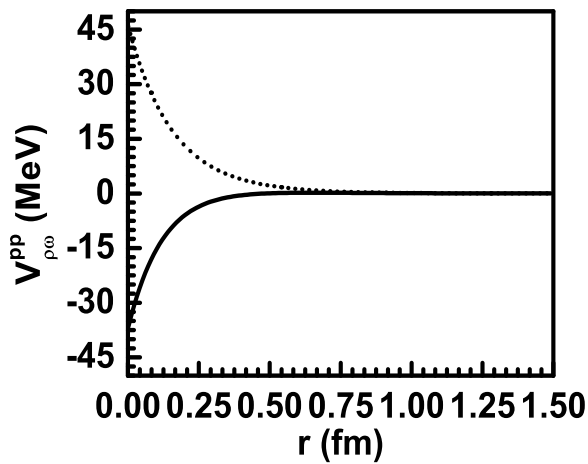}}
\caption{Total $^1S_0$ CSV potential with form factors (Eq.\ref{vrwf}) 
for $pp$ system is presented by the Solid curve and dotted 
curve shows the same without $\dt V^{pp}_{\rho\o}$.
\label{fig8}}
\end{center}
\end{figure}

The $^1S_0$ CSV potential with form factors is displayed in Fig.\ref{fig8}.
It is seen that the inclution of $\dt V^{NN}_{\rho\o}$ modifies the CSV potential
dramatically. It is to be noted that, with its inclusion, the CSV potential 
changes its sign.

The difference in scattering length $\Dt a$ when calculated with the potential 
of Eq.(\ref{vrwf}) is also markedly different from that calculated ignoring the 
term $\dt V^{NN}_{\rho\o}$. The results of $\Dt a$ with and without 
$\dt V^{NN}_{\rho\o}$ are shown in Table~\ref{dsc}.  

\begin{table}
\caption{The difference between $pp$ and $nn$ scattering lengths
at $^1S_0$.}
\begin{ruledtabular}
\label{dsc}
\begin{tabular}{lcc} 
         &  $\Dt a(C_\rho=0)$ (fm)   & $\Dt a(C_\rho=6.1)$ (fm) \\ \hline
  Without $\dt V^{NN}_{\rho\o}$ & $0.31$        & $2.14$    \\
  With $\dt V^{NN}_{\rho\o}$    & $-0.06$        & $-0.08$    \\
\end{tabular}
\end{ruledtabular}
\end{table}

\section{summary and discussion}

In the present work we have constructed the CSV potential within the 
framework of OBE model and studied the role of three momentum dependence 
$\rho^0$-$\o$ mixing amplitude in CSV. We find that the inclusion of the
contributions coming from the external legs are important because of the
strength of the $\rho NN$ tensor interactions. It is seen that unlike the 
previous finding \cite{Piekarewicz92} where the charge symmetry violation 
at the external legs were ignored, the strength of the CSV interaction could 
be significantly larger even when the off-shell amplitude for the $\rho^0$
-$\omega$ mixing is considered. It is important to note that contribution
from the spinors also modifies the central part of the two body potential
as shown in Eqs.(\ref{vcnn}) and (\ref{vrw}). Furthermore, we present results
both for the central and non-central part of the CSV potential. 

We also have calculated difference of $^1S_0$ scattering lengths between 
$pp$ and $nn$ systems and explicitly show the contribution of $\dt V^{NN}_{\rho\o}$. 
It is to be noted that $\Dt a$ changes sign with the inclusion of the fourier
transform of $\dt V^{NN}_{\rho\o}$. It would be interesting to apply the potential 
presented  here to calculate various other CSV observables to delineate the role 
of tensor interaction further.


\end{document}